# Computational Methodology for the Prediction of Functional Requirement Variations Across the Product Life-Cycle

**Guillaume Mandil[1], Alain Desrochers[1], Alain Rivière[2]**

[1]Université de Sherbrooke, Département de génie mécanique,
2500 boulevard de l'université, Sherbrooke Québec, Canada J1K 2R1

[2]LISMMA, 3 rue Fernand Hainaut, F-93407 Saint-Ouen Cedex, France

Guillaume.Mandil@Usherbrooke.ca

The great majority of engineered products are subject to thermo-mechanical loads which vary with the product environment during the various phases of its life-cycle (machining, assembly, intended service use…). Those load variations may result in different values of the parts nominal dimensions, which in turn generate corresponding variation of the effective clearance (functional requirement) in the assembly. Usually, and according to the contractual drawings, the parts are measured after the machining stage, whereas the interesting measurement values are the ones taken in service for they allow the prediction of clearance value under operating conditions. Unfortunately, measurement in operating conditions may not be practical to obtain. Hence, the main purpose of this research is to create, through computations and simulations, links between the values of the loads, dimensions and functional requirements during the successive phases of the life cycle of some given product.

The methodology presented is organised in three successive steps. Firstly, a functional requirement is chosen by the user, and the corresponding dimension chain is extracted from the Computer Aided Design (CAD) model. In order to be independent from the design parameters set by the designer, this paper uses the TTRS (Technologically and Topologically Related Surfaces) concept to relate the functional surfaces within a given dimension chain to some corresponding functional requirement at the manufacturing and assembly phase of the product life-cycle. Practically speaking, this leads to the definition of a set of nominal dimensions that serve as a baseline for the subsequent phases of the product life-cycle.

The second step consists in calculating the strains on the parts under thermo-mechanical loads in operating conditions. Generally this will be done using Finite Elements Analysis (FEA) or existing theoretical formulations. As this stage of the method uses existing techniques, the authors will use the simulation results as they are.

Thirdly, for each part of the product, the dimensions mentioned in the first step are adjusted with the results of the second step and introduced in the dimension chain. This, in turn, leads to a predictive value of the functional requirement under load. In the end, the complete methodology will provide the user with an account of the evolution of the functional requirement variation, across the main phases of the product life-cycle. Interestingly, these variations will add on top of the allowed manufacturing errors, as specified by the geometric dimensioning and tolerancing annotations from the initial design phase of the product life-cycle.

From an implementation point of view, the variations of the loads, temperature and dimensions will be expressed as intervals and will be associated to the parts and dimensions using attributes in the TTRS model. Furthermore, in order to be independent from the CAD software, the research will use STEP to represent 3D solids.

The paper concludes with the presentation of a practical application of the above methodology on a simple, one dimension crosshead guide example.

## 1 INTRODUCTION

Nowadays, design is increasingly widening in scope as additional views of a given product are being taken into account early in the product development process in order to lead to a better integration of marketing, engineering and costs requirements. This has translated as Design for X or DFX where X stands for Manufacturing (DFM), Assembly (DFA), etc. In a more comprehensive scope, Design should be embraced across the full life-cycle spectrum.

This would translate as additional product requirements reflecting the changing environment to which a product is being subjected throughout its life. In a more comprehensive way, a given functional requirement could have a different value in the life-cycle stage when it is useful compared to the stage when its value is being verified or measured. It then appears necessary to create a link between the different values of a given functional requirement to ensure theses values are compatible.

As each stage of the life-cycle occurs in a different set of environmental and utilisation conditions the mechanism is subject to thermo-mechanical load variations while stepping from one stage of the product life-cycle to another. As a consequence, an integrated design would take into account theses loads variations, hence helping the assembly meet a broader range of product requirements.

A typical application that illustrates best the above idea would be that of a jet engine for which the functional requirements varies during its own life-cycle. Indeed, the clearance between the rotor blades and engine housing (or stator) of the turbine will be quite different at assembly and in operation due to the high temperature and rotation velocity to which the rotor is subjected in service

The main purpose of this research is to create, through computations and simulations, links between the values of the loads, dimensions and functional requirements during the successive phases of the life cycle of some given product.

## 2 LITTERATURE REVIEW

As stated in the introduction, this work includes topics from three different fields. The current section mainly presents prior work in the field of functional requirements.

In this area many researches have been done on issues such as tolerance and dimension specification, tolerance analysis, tolerance synthesis, part geometry optimization, or geometry variations.

### 2.1 Related standards

In the GD&T[1] there are some standards. There are ANSI[2] standards edited by the ASME[3] [1-3]. These are specifying the semantic used to define geometrical features and their associated tolerances on 2D and 3D mechanical drawings. In addition the majority of the concepts related in theses standards have also been reproduced by the ISO[4] which has released a set of international standards. The organisation of this set [4] is synthetically presented by Marchèse [5]. Theses standards aspire to improve the consistence of geometric specifications with actual measuring techniques (Three-dimensional measuring machine, etc.) and to avoid ambiguity or any kind of user interpretation while using GD&T techniques. The benefits of using these techniques have been pointed out by Chiabert [6].

### 2.2 Related research

Here are presented prior research which has retained the authors' attention.

Firstly, Samper [7] presents an approach which allows considering the influence of both deformation of part and fit of joint into the analysis or synthesis of tolerances zones. The authors supposed that deformation of part and fit of joint have independent effects. A similar hypothesis which will be exposed below (in §4.1) is considered in this paper. In order to evaluate the influence of the two parameters named upper Samper uses four models to represent a mechanical assembly:

1. Rigid parts and perfect mechanism
2. Rigid parts and imperfect mechanism
3. Flexible parts and perfect mechanism
4. Flexible parts and imperfect mechanism

In this paper, calculations are made sequentially on the three firsts models, and the obtained results are compiled on the fourth which is the most complex. In a subsequent paper, the same authors [8] proposed an extension of their approach in order to make simultaneous calculations.

Secondly, Cid [9] presented a research which permits the evaluation of clearances under loads thanks to a clearance torsor introduced in [10]. This study investigates the case of the clearance of between a vehicle door and its frame. The representation of parts uses the simplification of considering 3D surfaces instead of 3D volumes.

---

[1] Geometric Dimensioning and Tolerancing
[2] American National Standards Institute
[3] American Society Of Mechanical Engineers
[4] International Organization for Standardization

## 3 DEFINITIONS AND CONCEPTS

Here are defined some terms used in the following sections of this paper. The example presented on figure 1 will be used to illustrate the following definitions.

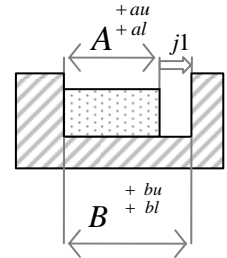

Figure 1 : Example for definitions

### 3.1 Dimensions

- Nominal dimension: effective dimension of the part used in the CAD model. This is noted $d_n$ in this section. In figure 1 there are two nominal dimensions graphically represented by $A$ and $B$.

- Tolerance: interval which define the acceptable variation of a measured dimension around its nominal value. This is noted *[tl;tu]* in this section. The boundaries of the tolerance interval (*tl* and *tu*) can be positive or negative. In figure 1, the tolerance associated with the dimension $A$ is *[al;au]*.

- Minimal dimension: minimal acceptable value for a measured dimension. This is noted $d_{min}$ in this section. $d_{min}= d_n+tl$. In figure 1, the tolerance minimal dimension of $A$ is $A\min = A + al$

- Maximal dimension: maximal acceptable value for a measured dimension. This is noted $d_{max}$ in this section. $d_{max}= d_n+tu$. In figure 1, the tolerance maximal dimension of $A$ is $A\max = A + au$

- Mean dimension: this is noted $\overline{d}$ in this section. $\overline{d} = d_n + (tl + tu)/2$. In the figure 1 example, the mean dimension of $A$ is $\overline{A} = A + (al + au)/2$.

- For each dimension defined above in this section $d_{(S1)}$ represents the dimension $d$ at the S1 stage of the life-cycle of the product.

### 3.2 Functional requirements

- Dimension chain: mathematical relation which links the value of a functional requirement with the dimensions of individual parts. In figure 1 the dimension chain associated with $j1$ functional requirement could be expressed thanks to equation (1) below where $i$ stand for whichever subscript. Two individual dimensions ($B$ and $A$) are involved in this chain.

$$j1_i = B_i - A_i \qquad (1)$$

- Nominal or real value of functional requirement: Value of a given functional requirement calculated with nominal or real dimensions. In figure 1 this could be expressed with $j1=B-A$ for nominal functional requirement and with $j1_r=B_r-A_r$ for real functional requirement

- Mean functional requirement: Value of a given functional requirement calculated with mean dimensions. In figure 1 this could be expressed thanks to expression (2) below.

$$\overline{j1} = (B + (bl + bu)/2) - (A + (al + au)/2) \qquad (2)$$

- Minimal and maximal functional requirement: theses values are calculated thanks to techniques of analysis of tolerance zones. Theses techniques stack-up tolerance zones specified for individual dimensions for a given functional requirement through the corresponding dimension chain. The result of this calculation is an interval which represents the possible range of variation for the functional requirement. This interval is centred on the mean value of the functional requirement. In the figure 1 example the following values are obtained: $j1\min = B\min - A\max$ and $j1\max = B\max - A\min$.

## 4 FUNCTIONAL REQUIREMENTS THROUGH LIFE-CYCLE STAGES

This section exposes how the evolution of the product along its life-cycle leads to changes of functional requirement values.

### 4.1 General Principles

Considering an assembly, there are two ways that the possible values of a given functional requirement vary.

There is first the stack-up of all the uncertainties due to machining and measuring techniques. This problem has been largely studied, and there exists some techniques of tolerances analysis to predict the possible variations of a given functional requirement. If the width of the possible range for a functional requirement $j_1$ is noted $\Delta j1$ then it's expressed thanks to: $\Delta j1 = j1\max - j1\min$. These variations are not directly linked to the whole product life-cycle. They are specifically related to the manufacturing and assembly stages. The manufacturing stage determines the accuracy of individual dimensions: the measured dimensions associated with their corresponding uncertainties have to meet the specified tolerances. Then, if required, part could be match at the assembly stage [11] in order to meet functional requirement specifications. In a more comprehensive way, the narrower the tolerances are, the smaller the range of possible variation for the functional requirement ($\Delta j1$) will be.

Secondly, the study must take into account the changing environment from an initial stage "Si" to a final stage "Sf" of the product life-cycle. One must notice that the terms initial and final are just relative to calculations steps. In a more comprehensive way the initial stage could be considered in operation and the final stage could be at the assembly situation. This allows expressing in the same formalism the shift from S1 to S2 stage and the shift from S2 to S1. Then, as mentioned in the introduction, the loads received by the assembly are subject to changes during this shift. This causes some deformation on the mechanism parts. Theses deformation can be viewed as a variation of individual dimensions. Let's precise which of the dimensions defined in section 3.1 are affected by part deformation.

If the parts of figure 1 are considered precise enough then it's possible to write down relation (3) below.

$$au - al << \overline{A} \tag{3}$$

Additionally, if the deformations are supposed to be small (i.e. $\Delta \overline{A} << \overline{A}$) then with (3) it's obtained the relation (4). This hypothesis is always verified if deformations are considered as elastic and linear which is the case in this research.

$$\Delta(au - al) << \Delta \overline{A} << \overline{A} \tag{4}$$

Equation (4) above means that the variation of the width of the tolerance zone is not, at a first order approximation, a significant source of variation compared to the variation of the mean value of the dimension. Consequently the width of the tolerance zone associated with the part at the stage "Si" is approximately the same as the width of the tolerance zone associated with the part at the stage "Sf".

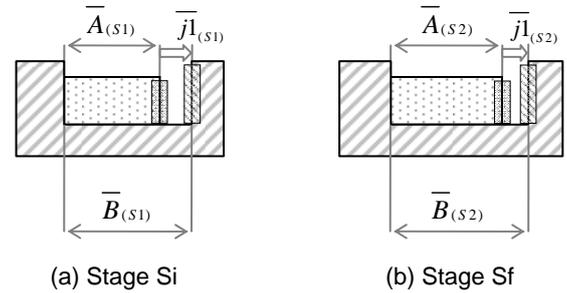

(a) Stage Si        (b) Stage Sf

Figure 2 : variation of individual dimensions due to loads

This explain that while stepping form "Si" to "Sf" life-cycle stage on figure 2 the width of the tolerance zone is unchanged and only the mean dimension is subject to some variations. This means that parts deformations due to loads can be viewed as a variation of mean values of individual dimensions. As tolerance stack-up techniques do not consider the dimensions but their possible variations one can infer that the value of $\Delta j_1$ do not vary across theses two life-cycle stages. Finally, variations of functional requirements across two life cycle stages can be represented on figure 3 below.

| Life-cycle stage | Value of Functional Requirement |
|---|---|
|  | −       0       + |
|  | Interference  \|  possible motion |
| Stage "Si" Stage "Sf" | $\Delta \overline{j1}$ / $\Delta j1$ / Mean value |

Figure 3 : Functional requirement variations due to loads

### 4.2 Computational rules

As exposed in the previous section, functional requirements are subject to some variations along the product life-cycle. In order to calculate the variation of the mean value of a given functional requirement several steps have to be followed. First, one has to calculate (using any existing technique) the deformation due to loads for each part. Then, for these deformations, corresponding mean values of the dimensions involved in dimension chain of the studied functional requirement have to be extracted. From there, it becomes possible to obtain the relation between the values of this functional requirement at two different stages of the product life-cycle. In accordance with previous section, from here, all values used in the following are means values.

These results have to be compared to the specifications for the functional requirements at the appropriate stage of the life cycle in order to ensure their continued compatibility while stepping along the life-cycle. Depending on the hypothesis and known variables three kinds of calculations can be done.

*From dimensions to functional requirement*

This computational approach (figure 4) is dimension driven meaning that changes on individual dimensions prompted by life-cycle evolution are translated into corresponding functional requirement values. This approach is used to check that dimensions chosen for the stage Si of the product life-cycle are compatible with required value of functional requirement at the stage Sf.

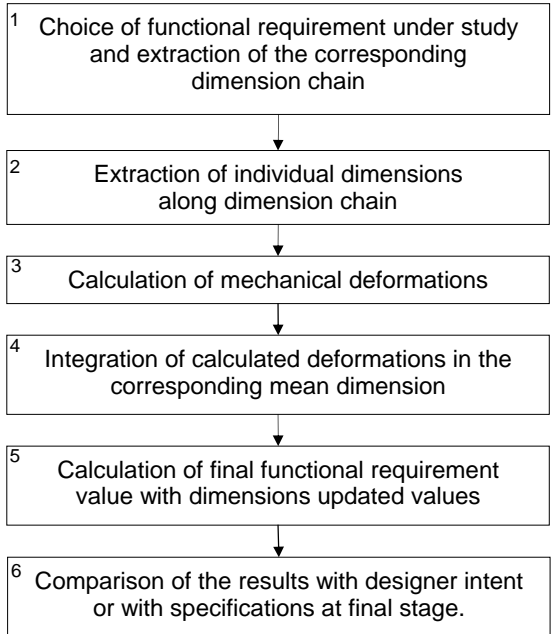

Figure 4 : Dimension driven calculation

In the figure 4 the steps 1 and 2 are not dependent on the product life-cycle. They are used to point out which functional requirement and dimension chain are under study. The dimension chain indicates the relation between the individual dimensions and the value of the functional requirement (§3.2). This relation is unchanged across the product life-cycle. Then, step 3 allows the shift from Si to Sf through calculations of the deformations due to loads variations. Finally steps 4 and 5 are required to express the results obtained at step 3 in terms of dimensions and functional requirement at the stage Sf of the product life-cycle. At step 6, this result is compared to specifications or to designer intent in order to validate the design of the product.

*From functional requirement to dimensions*

As opposed to the previous section, here the functional requirements are used as input to find compatible dimension values in the initial and final life-cycle stage. This calculation (figure 5) allow the designer to assign values to the individual dimensions of the product at the stage Si of the life-cycle in order to ensure the assembly meets a given value for the studied functional requirement under the Sf stage of the product life-cycle. As several dimensions are involved in the dimension chains, the result of this calculation is a range of acceptable values. In order to assign only one value to each dimension others criteria have to be considered.

In figure 5 the first step is life-cycle independent while it consists in choosing the functional requirement under study. The steps 2 and occurs in the initial stage of the product life-cycle. The result of these steps is the determination of initial individual dimensions values. Then, the calculations made at step 4 authorise the shift from Si to Sf. The results are expressed at the final stage in step 5. If necessary, the final value of the functional requirement could be calculated at step 6 thanks to individual dimension obtained at step 5. Finally these results are compared to specifications or to designer intent in order to validate the current design of the product.

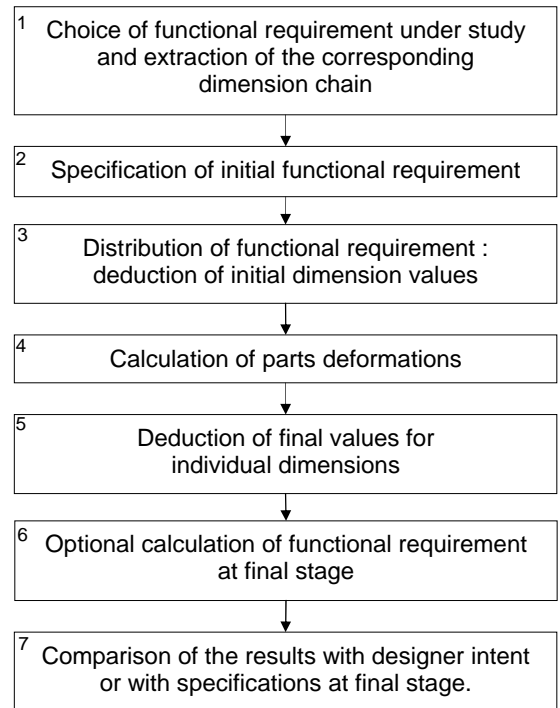

Figure 5 : Functional requirement driven calculation

*Determination of acceptable operating conditions*

Here it is assumed that the geometry of the product is completely defined. In a more comprehensive way, this means that the values of the functional requirements are specified both at initial and final stage of the product life-cycle. The aim of this calculation (figure 6) is to provide the designer with the acceptable variation of the environmental conditions and loads in relation with the specified values of a functional requirement along the product life-cycle.

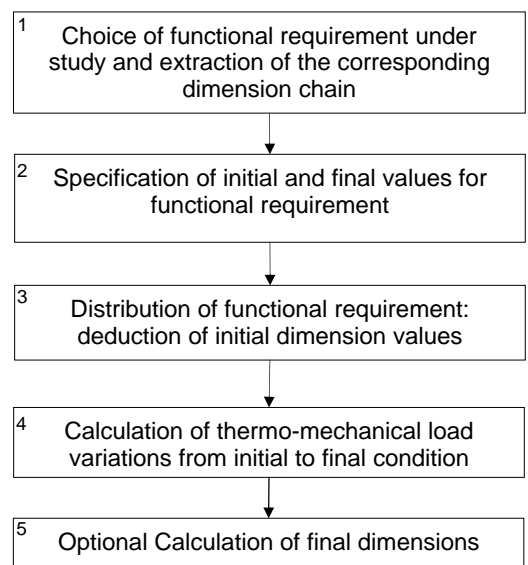

Figure 6 : Geometry driven calculation

The first three steps described in figure 6 establish the required input data for the problem at hand. Step 1 consists in choosing a functional requirement. Step 2 specifies its values at the initial and final stage of the life cycle. Afterwards, step 3 of the methodology assigns an initial value to individual dimensions involved in the chain. From there the dimensions at the initial phase of the product life cycle are obtained. Then the acceptable thermo-mechanical load variation is calculated at step 3. If several thermo-mechanical loads are subject to variations then the result of this step should be expressed as a range of possible variation for each thermo-mechanical load. Ultimately and if necessary, final dimensions can be calculated using the dimension driven technique previously exposed.

## 5 CASE STUDY: A SIMPLE 1D CROSSHEAD GUIDE

In this section a simple application case is presented. The studied guide presented on figure 7 is constituted of a one-piece wheel shaft positioned in a one-piece frame.

The authors deliberately choose a simple example in order to present in a simple way the benefits and the perspectives of their approach. This case is considered as one-dimensional and the thermo-mechanical loads are limited to temperature variation. It's admitted that the chosen case is not realistic, but it's simple enough to be calculated and verified with existing theoretical formulations.

### 5.1 Hypothesis

*Life-Cycle*

This case will be studied while stepping from the initial stage Si to the final stage Sf of this product life-cycle. It should be noted however those generic initial and final life-cycle stages do not necessarily have to follow a temporal sequence. In other words, depending on the problem perspective, the initial stage could be the product use in operation whereas the final stage would be the product at the assembly phase.

*Geometry*

The geometry chosen for this study is presented on figure 7 below. It's described thanks to six dimensions.

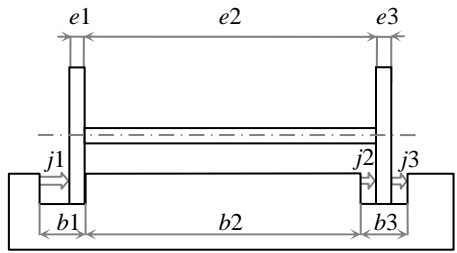

Figure 7 : 1D case study

For all the following calculations the dimensions $e1$ $e2$ and $e3$ are considered as hard constraints and their imposed values at 20°C are exposed in table 1 below. The example case study consists in designing the frame.

| Dimension | Value at 20°C |
|---|---|
| $e1_{(20°C)}$ | $60^{\pm 0,1}\, mm$ |
| $e2_{(20°C)}$ | $1440^{\pm 0,1}\, mm$ |
| $e3_{(20°C)}$ | $60^{\pm 0,1}\, mm$ |

Table 1 : Dimension of the shaft at 20°C

*Materials*

The wheel shaft is built in aluminium and the frame is made of steel.

*Loads and behaviour law*

It's supposed that during its life-cycle this mechanism is subject to a temperature variation which will result in linear deformations such as presented in equation (5) below.

$$l_{(Sf)} - l_{(Si)} = \alpha \cdot l_{(Si)} \cdot \left(t_{(Sf)} - t_{(Si)}\right) \quad (5)$$

In equation (5) $l$ represents the length of a given dimension and $t$ stands for the temperature. Finally, $\alpha$ designates the coefficient of thermal dilatation. Typical values of $\alpha$ are presented in table 2 below.

| Material | Notation | Coefficient of thermal dilatation |
|---|---|---|
| Steel | $\alpha_s$ | $1.20\ E\text{-}05\ K^{-1}$ |
| Aluminium | $\alpha_a$ | $2.38\ E\text{-}05\ K^{-1}$ |

Table 2 : Typical values of coefficient of thermal dilatation

*Functional requirements*

This case also defines three functional requirements $j1$ $j2$ and $j3$ which will be studied. It's also supposed that each functional requirement can have a minimum and/or a maximum required value for each stage of the product life cycle.

*Dimension chains*

For each requirement it's possible to define a dimension chain from which the following equations (6)(7)(8) are derived to calculate the values of the functional requirements.

$$\begin{cases} j1 = b1 - e1 \\ j1\min = b1\min - e1\max \\ j1\max = b1\max - e1\min \end{cases} \quad (6)$$

$$\begin{cases} j2 = e2 - b2 \\ j2\min = e2\min - b2\max \\ j2\max = e2\max - b2\min \end{cases} \quad (7)$$

$$\begin{cases} j3 = b2 + b3 - e2 - e3 \\ j3\min = b3\min - e3\max - e2\max + b2\min \\ j3\max = b3\max - e3\min - e2\min + b2\max \end{cases} \quad (8)$$

*Implementation*

All the above exposed hypothesis have been introduced in an Excel spreadsheet for each kind of calculation presented in §4.2. For these calculations two life-cycle stages Si and Sf are considered.

### 5.2 Dimension driven calculation

This first calculation aims at answering the question: "What will be the value of a given functional requirement after thermal dilatation of the parts?"

*Hypothesis*

The studied functional requirements are $j1$ $j2$ and $j3$ which can be calculated thanks to the dimension chains (6)(7)(8).

| Variable | | Value |
|---|---|---|
| $t_{(Si)}$ | | $20°C$ |
| $t_{(Sf)}$ | | $50°C$ |
| $b1_{(Si)}$ | at 20°C | $60.3^{\pm 0,1}\, mm$ |
| $b2_{(Si)}$ | at 20°C | $1439.7^{\pm 0,1}\, mm$ |
| $b3_{(Si)}$ | at 20°C | $60.8^{\pm 0,1}\, mm$ |

Table 3 : first calculation hypothesis

This represents the steps 1 and 2 of the figure 4 methodology. As the initial temperature is 20°C the values of $e1$ $e2$ and $e3$ are those presented in table 1. The initial values presented in table 1 and table 3 will be introduced in the Excel spreadsheet in order to map the value of functional requirements along with the temperature variations.

*Calculations*

This is the 3$^{rd}$ and 4$^{th}$ step of figure 4 methodology. The above exposed hypotheses are then used as input for calculating the deformations thanks to equation (5). The obtained results are presented in table 4.

| Initial dimension at stage Si (at 20°C) | Deformation | Final dimension at stage Sf (50°C) |
|---|---|---|
| $\overline{e1_{(Si)}} = 60mm$ | $\Delta\overline{e1} = 0.043mm$ | $\overline{e1_{(Sf)}} = 60.043mm$ |
| $\overline{e2_{(Si)}} = 1440mm$ | $\Delta\overline{e2} = 1.028mm$ | $\overline{e2_{(Sf)}} = 1441.028mm$ |
| $\overline{e3_{(Si)}} = 60mm$ | $\Delta\overline{e3} = 0.043mm$ | $\overline{e3_{(Sf)}} = 60.043mm$ |
| $\overline{b1_{(Si)}} = 60.3mm$ | $\Delta\overline{b1} = 0.022mm$ | $\overline{b1_{(Sf)}} = 60.322mm$ |
| $\overline{b2_{(Si)}} = 1439.7mm$ | $\Delta\overline{b2} = 0.518mm$ | $\overline{b2_{(Sf)}} = 1440.218mm$ |
| $\overline{b3_{(Si)}} = 60.8mm$ | $\Delta\overline{b3} = 0.022mm$ | $\overline{b3_{(Sf)}} = 60.822mm$ |

Table 4 : Deformation and final dimensions for dimension driven calculation

*Results*

For the 5$^{th}$ step of figure 4 methodology the values presented in table 4 are introduced in dimensions chains (6)(7)(8). The results for the functional requirements are presented in the table 5 below.

| Functional requirement | | Mean values | Range of possible values |
|---|---|---|---|
| $j1_{(Si)}$ | at 20°C | 0.3 mm | [0.1 ; 0.5] mm |
| $j2_{(Si)}$ | at 20°C | 0.3 mm | [0.1 ; 0.5] mm |
| $j3_{(Si)}$ | at 20°C | 0.5 mm | [0.1 ; 0.9] mm |
| $j1_{(Sf)}$ | at 50°C | 0.279 mm | [0.079 ; 0.479] mm |
| $j2_{(Sf)}$ | at 50°C | 0.810 mm | [0.610 ; 1.010] mm |
| $j3_{(Sf)}$ | at 50°C | -0.031 mm | [-0.431 ; 0.369] mm |

Table 5 : Results of dimension driven calculation

First, the tolerances along dimension chains (6)(7)(8) have been analysed in order to calculate the range of possible values for the functional requirements at stage Si. Then, mean values of functional requirement are calculated at stage Sf and they are associated with the range of possible variations resulting from the tolerance analysis and which do not vary along the life-cycle (cf. § 4.1).

*Conclusion*

The results of table 5 show that at the stage Sf (under 50°C) there might appear some interference on $j3$ functional requirement. If this interference is not compatible with the product functionality (step 6 in figure 4) then the dimensions of the mechanism must be reviewed.

**5.3 Functional requirement driven calculation**

"Which dimension has to be chosen in order to obtain a given value of a functional requirement after thermal dilatation?" In this section the studied problem is the calculation of the dimensions of the frame at 20°C which ensure given values for functional requirements $j1$ $j2$ and $j3$ at 50°C (step 1 of the figure 5 methodology).

*Hypothesis*

Considering the objectives exposed above, the initial conditions are considered at 50°C and the final stage is at 20°C (cf. table 7). The targeted values for functional requirement are shown in table 6 (steps 1 and 2 in figure 5).

| Functional requirement | | Mean values | Acceptable values |
|---|---|---|---|
| $j1_{(Si)}$ | at 50°C | 0.25 mm | [0.05 ; 0.45] mm |
| $j2_{(Si)}$ | at 50°C | 0.4 mm | [0.2 ; 0.6] mm |
| $j3_{(Si)}$ | at 50°C | 0.45 mm | [0.05 ; 0.85] mm |

Table 6 : values for functional requirement at 50°C

Moreover, the dimensions of the shaft at 50°C can be found in table 4. Therefore, the values of the dimensions and tolerance zones of the frame at 50°C can be deduced thanks to dimension chains (6)(7)(8) (step 3 in figure 5). These values are exposed in table 7.

| Variable | | Value |
|---|---|---|
| $t_{(Si)}$ | | 50°C |
| $t_{(Sf)}$ | | 20°C |
| $e1_{(Si)}$ | at 50°C | $60.043^{\pm 0.1}$ mm |
| $e2_{(Si)}$ | at 50°C | $1441.028^{\pm 0.1}$ mm |
| $e3_{(Si)}$ | at 50°C | $60.043^{\pm 0.1}$ mm |
| $b1_{(Si)}$ | at 50°C | $60.293^{\pm 0.1}$ mm |
| $b2_{(Si)}$ | at 50°C | $1440.628^{\pm 0.1}$ mm |
| $b3_{(Si)}$ | at 50°C | $60.893^{\pm 0.1}$ mm |

Table 7 : Functional requirement driven calculation inputs

*Calculations*

The calculations of thermal deformation (step 4 in figure 5) are performed in the same way as in section 5.2.

*Results*

Final dimensions after step 5 in figure 5 are presented in table 8.

| Final dimension at stage Sf (20°C) |
|---|
| $\overline{e1_{(Sf)}} = 60mm$ |
| $\overline{e2_{(Sf)}} = 1440mm$ |
| $\overline{e3_{(Sf)}} = 60mm$ |
| $\overline{b1_{(Sf)}} = 60.271mm$ |
| $\overline{b2_{(Sf)}} = 1440.109mm$ |
| $\overline{b3_{(Sf)}} = 60.871mm$ |

Table 8 : Final dimensions at 20°C for functional requirement driven calculation

The optional 6$^{th}$ step of figure 5 is used there as a validation process for individual dimensions. The calculations are made in the same way as in section 5.2. Final values of functional requirements are presented in table 9 below.

| Functional requirement | | Mean values | Acceptable values |
|---|---|---|---|
| $j1_{(Sf)}$ | at 20°C | 0.271 mm | [0.071 ; 0.471] mm |
| $j2_{(Sf)}$ | at 20°C | -0.110 mm | [-0.310 ; 0.09] mm |
| $j3_{(Sf)}$ | at 20°C | 0.981 mm | [0.581 ; 1.381] mm |

Table 9 : values for functional requirement at 20°C

*Conclusion*

The results of table 9 show that at the stage Sf (under 20°C) there might appear some interference on *j*2 functional requirement. If this interference is not compatible with the product functionality (step 7 in figure 5) then the dimensions of the mechanism must be reviewed. A greater initial mean value for the functional requirement or a narrower range for the corresponding tolerance zone must be used as an updated input for this calculation.

### 5.4 Geometry driven calculation

"Which loads are acceptable in order to ensure the respect of a functional requirement at two stages Si and Sf of the life-cycle?"

*Hypothesis*

First, the designer must impose a mean value to functional requirements at stages Si and Sf of the product life-cycle. Initial temperature is set to 20°C. Thanks to table 1 dimensions, dimension chain expressions (6)(7)(8) and functional requirement values at Si one can deduce values of $\overline{b1_{(Si)}}$ $\overline{b2_{(Si)}}$ and $\overline{b3_{(Si)}}$. These values can be found in table 10 below.

| Variable | | Value |
|---|---|---|
| $t_{(Si)}$ | | 20 °C |
| $\overline{j1_{(Si)}}$ | at 20°C | 0.3 mm |
| $\overline{j2_{(Si)}}$ | at 20°C | 0.3 mm |
| $\overline{j3_{(Si)}}$ | at 20°C | 0.5 mm |
| $\overline{j1_{(Sf)}}$ | | 0.25 mm |
| $\overline{j2_{(Sf)}}$ | | 0.4 mm |
| $\overline{j3_{(Sf)}}$ | | 0.45 mm |
| $e1_{(Si)}$ | at 20°C | $60^{\pm 0,1}$ mm |
| $e2_{(Si)}$ | at 20°C | $1440^{\pm 0,1}$ mm |
| $e3_{(Si)}$ | at 20°C | $60^{\pm 0,1}$ mm |
| $b1_{(Si)}$ | at 20°C | $60.3^{\pm 0,1}$ mm |
| $b2_{(Si)}$ | at 20°C | $1439.7^{\pm 0,1}$ mm |
| $b3_{(Si)}$ | at 20°C | $60.8^{\pm 0,1}$ mm |

Table 10 : Geometry driven calculation inputs

This represents steps 1 and 2 of figure 6 methodology.

*Calculations*

This part of the method represents the 3rd step of figure 6 methodology. The use of dimension chains and deformation law will be detailed here for the calculation relative to j3 functional requirement. A similar approach is used for *j*1 and *j*2. First, the expression of the dimension chain (8) is used to calculate the mean value of the functional requirement j3 given at initial stage (9) and at final stage (10).

$$\overline{j3_{(Si)}} = \overline{b2_{(Si)}} + \overline{b3_{(Si)}} - \overline{e2_{(Si)}} - \overline{e3_{(Si)}} \quad (9)$$

$$\overline{j3_{(Sf)}} = \overline{b2_{(Sf)}} + \overline{b3_{(Sf)}} - \overline{e2_{(Sf)}} - \overline{e3_{(Sf)}} \quad (10)$$

From these expressions the variations of these mean values are deduced with (11).

$$\Delta \overline{j3} = \overline{b2_{(Sf)}} - \overline{b2_{(Si)}} + \overline{b3_{(Sf)}} - \overline{b3_{(Si)}} - \overline{e2_{(Sf)}} + \overline{e2_{(Si)}} - \overline{e3_{(Sf)}} + \overline{e3_{(Si)}}$$
$$\text{with } \Delta \overline{j3} = \overline{j3_{(Sf)}} - \overline{j3_{(Si)}} \quad (11)$$

Additionally, with behaviour law (5), the following set (12) of life-cycle dependence relation is obtained.

$$\begin{cases} e3_{(Sf)} = \alpha_a \cdot e3_{(Si)} \cdot (t_{(Sf)} - t_{(Si)}) + e3_{(Si)} \\ e2_{(Sf)} = \alpha_a \cdot e2_{(Si)} \cdot (t_{(Sf)} - t_{(Si)}) + e2_{(Si)} \\ b3_{(Sf)} = \alpha_s \cdot b3_{(Si)} \cdot (t_{(Sf)} - t_{(Si)}) + b3_{(Si)} \\ b2_{(Sf)} = \alpha_s \cdot b2_{(Si)} \cdot (t_{(Sf)} - t_{(Si)}) + b2_{(Si)} \end{cases} \quad (12)$$

Finally, with the substitution of final values (12) in equation (11) the equation (13) gives the expression of the admissible final temperature for *j*3.

$$t_{(Sf)} = t_{(Si)} + \frac{\Delta \overline{j3}}{\alpha_s \cdot \left(\overline{b3_{(Si)}} + \overline{b2_{(Si)}}\right) - \alpha_a \cdot \left(\overline{e3_{(Si)}} + \overline{e2_{(Si)}}\right)} \quad (13)$$

*Results*

The calculations presented in the previous paragraph give the results presented in the following table 11.

| Functional requirement to be respected | Admissible temperature at final stage |
|---|---|
| j1 | 91.0 °C |
| j2 | 25.9 °C |
| j3 | 22.8 °C |

Table 11 : Admissible temperature for geometry driven calculation

These results are those obtained after the 4th step of figure 6 methodology. A 5th step consists in choosing the most restrictive value for the final temperature which is 22.8 °C in this case. Under this condition, final dimensions can be deduced thanks to a dimension driven calculation (§5.2).

*Conclusion*

This resulting limited range of temperature variation means that if solutions presented in the conclusion of section 5.3 are not applicable, then, the only way to ensure mechanism functionality is to reduce the possible range of load variation.

## 6 CONCLUSION AND PERSPECTIVES

### 6.1 High level management of functional requirement

This simple study has shown the interest of considering functional requirement variations along the various phases of the product life cycle. As the proposed approach does not affect the width of the tolerance zones, it consequently does not impact machining costs either. This aspect is very interesting because it allows possible improvements of the mechanism functionality at a given stage of the product life-cycle without increasing its cost. Moreover, the previous section (§5) has been organized as a high-level design methodology. This methodology consists in three successive steps. First, the current design of the product is checked with a dimension driven calculation (§5.2). If the resulting functional requirement is not meeting products specifications at the target of final phase of the product life-cycle, then a second step of redesign is done using a functional requirement driven calculation (§5.3). The initial stage of the product life-cycle for the functional requirement driven calculation should correspond to the final stage of the dimension driven one. For example, if the dimension driven calculation is made from the assembly stage to the operation stage of the product life-cycle, then the functional driven calculation will be done from operation to assembly in order to assign individual dimensions at the assembly stage of the product life cycle (which are then considered as the design variables). Finally, if this fails and acceptable values for individual dimension can not be found, an ultimate geometry driven calculation (§5.4) can be used to compute the range of acceptable thermo-mechanical loads variation between the two stages considered in the product life-cycle.

**6.2 Further work**

This study has been illustrated on a very restrictive case and must be extended to a more general case. First this model should be able to tackle 2D and 3D geometries. This must include an appropriate way for the mathematical representation of 2D and 3D dimension chains and an appropriate tool for the calculation of parts deformations under thermo-mechanical loads. The preliminary studies done in this way suggest the use of TTRS[5] and MGRE[6] [12] for the representation of dimension chains. Additionally, Finite Element Analysis (FEA) appears as the most suitable tool for the calculation of parts deformation as it is commonly used in the mechanical industry. Worth noting is the fact that 2D and 3D dimension chains contain both linear and angular dimensions which must be accounted for by the proposed approach.

Concerning the life-cycle aspects of the methodology, some improvements are planned. There is first the combination of several thermo-mechanical loads as a source of geometrical variations. Secondly the possibility of further constraining the design by specifying values for the functional requirements at various stages of the product life-cycle will be investigated.

**7 ACKNOWLEDGMENTS**

The authors wish to acknowledge the financial support of the Natural Science and Engineering Council of Canada for funding this research through its Discovery grant program.

---

[5] Technologically and Topologically Related Surfaces

[6] Minimum Geometric Reference Element